\documentclass[10pt,letterpaper]{article}

\usepackage{url,hyperref,lineno,microtype,subcaption}
\usepackage[onehalfspacing]{setspace}
\usepackage{graphicx}

\begin{document}

\begin{flushleft}
{\Large
\textbf\newline{Evolution of Human-like Social Grooming Strategies regarding Richness and Group Size} 
}
\newline
\\
Masanori Takano\textsuperscript{1},
Genki Ichinose\textsuperscript{2}
\\
\bigskip
\textbf{1} Akihabara Laboratory, CyberAgent, Inc., Tokyo, Japan \\
\textbf{2} Department of Mathematical and Systems Engineering, Shizuoka University, Hamamatsu, 432-8561, Japan
\\
\bigskip

* takano\_masanori@cyberagent.co.jp.

\end{flushleft}

\section*{Abstract}

Human social strategies have evolved as an adaption to behave in complex societies.
In such societies, humans intensively tend to cooperate with their closer friends, because they have to distribute their limited resources through cooperation (e.g. time, food, etc.).
It also makes the situation difficult to have uniform social relationships (social grooming) with all friends.
Thus, the social relationship strengths often show a much skewed distribution (a power law distribution).
Here we aim to show adaptivity of such social grooming strategies in order to explore the evolution of human social intelligence.
We use a model in the framework of evolutionary games where the social grooming strategies evolve via building social relationships with cooperators.
Simulation results demonstrate four evolutionary trends.
One of the trends is similar to the strategy that humans use.
We find that these trends depend on three parameters; individuals' richness, group sizes, and the amount of social grooming.
The human-like strategy evolves in large poor groups.
Moreover, the increase of the amount of social grooming makes the group size larger.
Conversely, this implies that the same strategy evolves when the amount of social grooming is properly adjusted even if the group sizes are different.
Our results are important in the sense that, between human and non-human primates, the differences of the group size and the amount of social grooming are significant.


\section{Introduction}

Cooperation is common among humans and it is fundamental to our society~\cite{Fehr2003,Smitha}.
The amount of cooperation by other people is limited because they have to pay costs (e.g., money, time, opportunities, food, etc.)~\cite{Santos2006a,Xu2015}.
Therefore, people carefully choose their friends in order to receive intensive cooperation~\cite{Rand2011,Grujic2012,Wang2012}.

Actually, people tend to cooperate with close friends.
An experimental study using the Donation Game shows that participants tend to cooperate more with closer friends~\cite{Harrison2011}.
Another study using the Public Goods Game shows that friend groups are more cooperative with each other than with other groups~\cite{haan2006}.
Additionally, in a data analysis study dealing with the data set of a social network game, people's frequent communication increases their cooperative behavior~\cite{takano_ngc,takano_socinfo}.

Thus, it is important that humans have stronger social relationships in greater numbers with cooperators than with others.
We define social grooming as the behaviour that constructs social relationships.
Primarily, social grooming is the act of cleaning or maintaining the body of a social partner in primates~\cite{Dunbar,Dunbar2000,Nakamura2003}.
Social bonding is part of the functional aspect of social grooming.
Therefore, human social bonding behavior is also called social grooming~\cite{Dunbar,Dunbar2000}, as a hypothetical extrapolation of the findings in non-human animals.

The behavior constructing social relationships is not limited to humans but widely observed in primates~\cite{Kobayashi1997,Kobayashi2011,Dunbar,Dunbar2000,Nakamura2003,takano_ngc,takano_socinfo,takano2016_palg}.
In doing so, they face cognitive constraints~\cite{Dunbar2012} (e.g. memory and processing capacity) and time constraints (i.e. time costs) in constructing and maintaining social relationships.
These time constraints are not negligible, as people spend a fifth of their day in social grooming~\cite{Dunbar1998a} for maintaining the  relationship~\cite{Hill2003,ROBERTS2011}.
Therefore, the strength of existing social relationships exhibits a negative correlation with the total number of social relationships~\cite{Roberts2009, Miritello2013}.

On the other hand, it is important to select cooperative partners in the evolution of cooperation because cooperators tend to be exploited by defectors~\cite{Axelrod}.
Moreover, direct reciprocity, spatial reciprocity, and network reciprocity, which are the mechanisms for the evolution of cooperation, suggest the necessity of fixed relationships~\cite{Perc2010,Perc2017}.
Therefore, it is reasonable to consider that humans and other social animals tend to cooperate with their close partners~\cite{Harrison2011,haan2006,takano_ngc,takano_socinfo}.



Humans must construct and maintain social relationships within the constraints of this trade-off.
We expect that strategies are employed to distribute the limited time resources to maximize benefits from their social relationships~\cite{Brown2006,Miritello2013a,Saramaki2014}.
As a result of such strategies, social relationship strengths, as measured by frequency of social grooming~\cite{ROBERTS2011,Arnaboldi2012,Song2012,Arnaboldi2013a,Fujihara2014,Saramaki2014,takano2016_palg}, may often show a skewed distribution~\cite{Zhou2005,Arnaboldi2013a} (distributions following a power law~\cite{Song2012,pachur2012,Arnaboldi2012,Fujihara2014,Hossmann2011,Timm-Giel2013,takano2016_palg}).
Moreover, it has been demonstrated that social structures of non-human primates~\cite{Dunbar2012a,Kanngiesser2011,Tung2015,leve2016} are also skewed.

The skewed distributions of the relationships could be generated by a strategy where individuals select social grooming partners in proportion to the strength of their social relationships~\cite{pachur2012,takano2016_palg}; known as the Yule--Simon process~\cite{Yule1925,SIMON1955,Newman2005}.
Individuals should pay time costs to win the competitions with others by strengthening their social relationships with cooperators, assuming that having strong social relationships is to receive cooperation.

Human societies using these strategies are much larger than those of non-human primates.
Based on the social brain hypothesis, human intelligence has evolved to adapt to large societies.
Therefore, the evolution of human strategies of social relationship construction may explain the origin of human intelligence.
However, evolutionary stability of the strategies, i.e. the Yule--Simon process, is still open investigation.

In this paper, we aim to show the adaptivity of the social grooming strategies in order to explore the evolution of human social intelligence predicted by the social brain hypothesis.
Especially, we focus on how environments drive the evolution of a social grooming strategy that humans use in their daily life.
The evolution should depend on group size and the amount of resources for cooperation.
For this purpose, we simulate the evolution of the strategy to receive cooperation from others with different environmental conditions for cooperations.
We show that strategies evolve depending on the strength of social relationships.

\section{Methods}

\begin{figure}[h!]
\begin{center}
\includegraphics[width=0.95\linewidth]{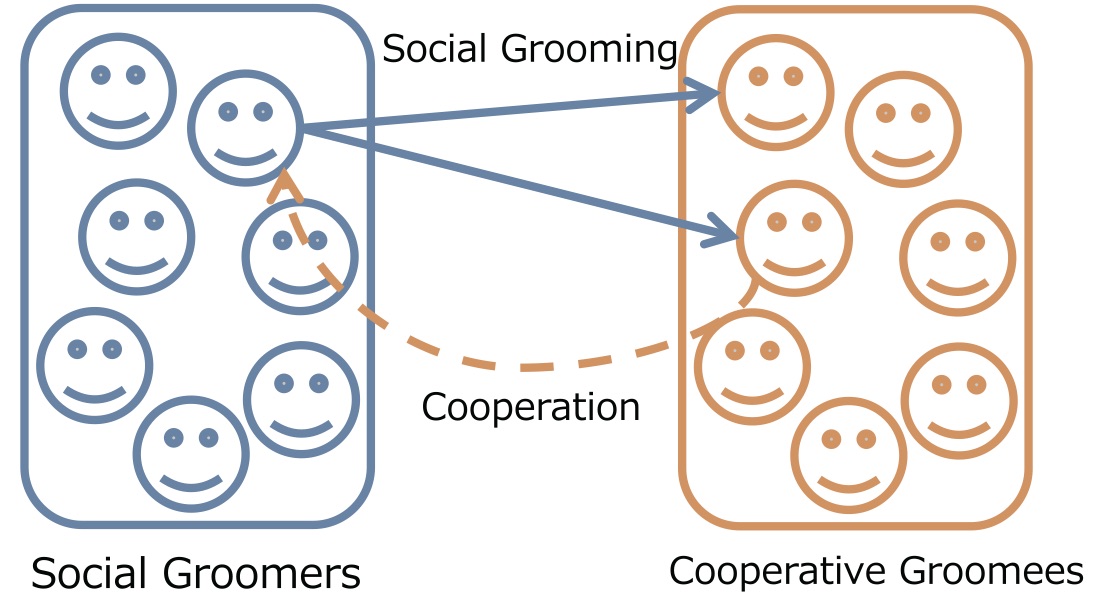}
\caption{
Concept of our model.
Social groomers interact with cooperative groomees depending on their social grooming strategies.
Cooperative groomees cooperate with social groomers who are top $R_c$ on the strengths of social relationships.
Groomer strategies evolve based on their fitness which is the amount of cooperation from groomees.
}
\label{fig_abst}
\end{center}
\end{figure}

We expand the model of \cite{takano2016_palg} to an evolutionary game.
They consider two types of individuals; social groomers and cooperative groomees (Fig.~\ref{fig_abst})~\cite{takano2016_palg}.
In the real world, individuals are groomers and groomees, simultaneously.
For simplicity, they use this classification to focus on the social grooming strategies for social structures.
In this paper, we focus on the evolution of social grooming strategies.
While the evolutionary dynamics of cooperation are well known~\cite{Nowak2006,Perc2010,Rand2013,Perc2017}, there are few study on the evolutionary dynamics of social grooming.
Groomers construct their social relationships with groomees depending on their social grooming strategies in a ``grooming stage.''
Cooperative groomees cooperate with groomers depending on social relationship strengths in a ``cooperation stage.''
Groomer strategies evolve based on their fitness which is the amount of cooperation from groomees in each generation.
Groomees' cooperation strategies are static.

In a grooming stage, groomer $i$ repeatedly interacts with cooperative groomees $R_g$ times depending on their social grooming strategy $(s_i, q_i)$.
$q_i$ is a ratio that $i$ constructs a new social relationship with a stranger, new groomee $j$, and $s_i$ is a parameter of a probabilistic function $p(d_{ij}; s_i)$ which selects existing social grooming partner $j$ depending on $d_{ij}$ ($d_{ij}>0$).
We used the following function (Fig.~\ref{fig_str_sample}) as a simple function to express various strategies depending on $d_{ij}$ including concentrated investment to strong relationships ($s=4$), diversified investment to weak relationships ($s=-4$), at random ($s=0$), and the Yule--Simon process ($s=1$; i.e. human-like strategy).

\begin{eqnarray}
p(d_{ij};s_i)=b(d_{ij};\alpha_i, \beta_i)/\sum_{k=1}^M b(d_{ik};\alpha_i, \beta_i),
\label{eq_sel}
\end{eqnarray}
where $\alpha_i=1+s_i, \beta_i=1$ when $s_i \geq 0$ while $\alpha_i=1, \beta_i=1-s_i$ when $s_i < 0$.
$d_{ij}$ is $w_{ij}/max(\{w_{i1}, w_{i2}, \dots, w_{iM}\})$, where $w_{ij}$ shows strength of social relationships, i.e., the number of social grooming from $i$ to $j$.
This function only depends on $d_{ij}$, because previous studies have revealed that people select their social grooming partners depending on the strength of social relationships~\cite{pachur2012,takano2016_palg}.
Therefore, this function can simply represent human-like social grooming strategies.
$M$ is the number of groomees.
$b(x; \alpha, \beta)$ is a normalized beta distribution $x^{\alpha-1}(1-x)^{\beta-1} / B(\alpha, \beta)$, where $B(\cdot, \cdot)$ is a beta function.

In a cooperation stage, groomee $j$ cooperates with groomers in the top $R_c$ as ranked by $\{w_{1j}, w_{2j}, \dots, w_{Nj}\}$.
The total payoff (i.e. fitness) of each groomer is the number of cooperation (i.e. the number of times ranked in the top $R_c$ of each cooperator).
That is, cooperators cooperate in their close relationships according to their resources $R_c$.
$R_c M$ shows all resources in the environment $(R_c, M)$, i.e. the total amount of cooperation.

The next generation is generated by sampling with replacement in proportion to the groomers' fitness, i.e. the roulette wheel selection.
In each generation, $s$ mutates by the Gaussian distribution ($\mu=0, \sigma=0.2$) and $q$ mutates by the Gaussian distribution ($\mu=0, \sigma=0.05$), where $\mu$ is a mean and $\sigma$ is a standard deviation of the distribution, where $q \in [0, 1]$ (if $q$ is out of range by mutation, then it is set to the nearest value in $0$ or $1$).
Groomers' $s$ and $q$ in an initial generation are set by the Gaussian distribution ($\mu=0, \sigma=5.0$) and by uniform distribution $[0, 1]$, respectively.
Cooperators do not evolve.

We conducted evolutionary simulations 30 times on each $R_c$ and $M$ by using this model ($R_c \in \{5, 10, \dots, 50\}$, $M \in \{5, 10, \dots 200\}$).
The number of groomers $N$ is $100$, the number of social grooming actions $R_g$ in each grooming stage is $300$ (we also use $R_g=100$ in experiments), and the number of generation $T$ is $200$.
The source code is available at ``https://doi.org/10.6084/m9.figshare.5526850.v1''.

\section{Results}

\begin{figure}[h!]
\begin{center}
\includegraphics[width=0.95\linewidth]{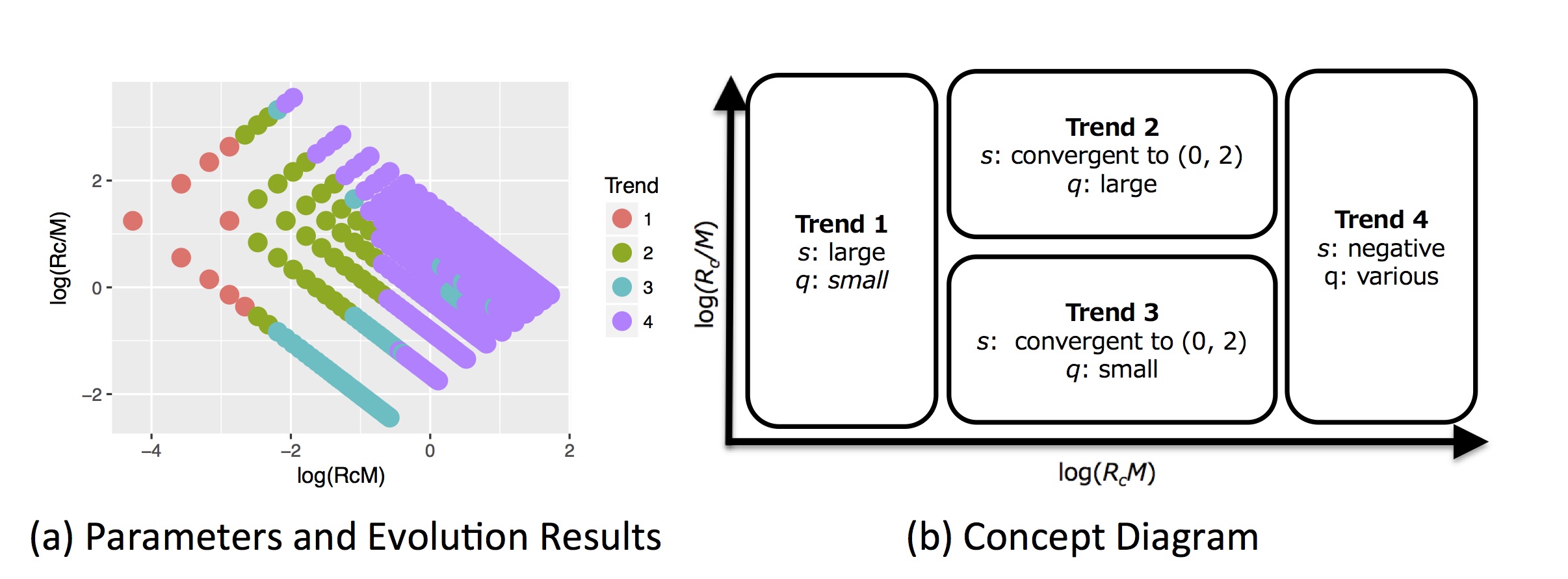}
\caption{
Summary of results of evolutionary simulations.
We found four evolutionary trends ($s$ and $q$ of the final populations) depending on total resources $R_c M$ and the ratio of each cooperator's resources to the number of cooperators $R_c/M$ (see details Fig.~\ref{fig_map} and Fig.~\ref{fig_map100}).
Fig.~a shows the results of evolution with parameter $R_c$ and $M$.
Each color shows the most frequent trend in parameters of the point.
This was created based on Fig.~\ref{fig_map}b.
Fig.~b is the concept diagram.
Trend 1 evolved when $R_c M$ was small.
Trend 4 evolved when $R_c M$ was large.
Trends 2 and 3 evolved in the intermediate range between trends 1 and 4 where $R_c/M$ determined whether groomers evolved to trend 2 or 3.
The behavior of trends 2 and 3 were similar to human strategies, although trend 2 was closer, as described.
}
\label{fig_smr}
\end{center}
\end{figure}

\begin{figure}[h!]
\begin{center}
  \includegraphics[width=0.95\linewidth]{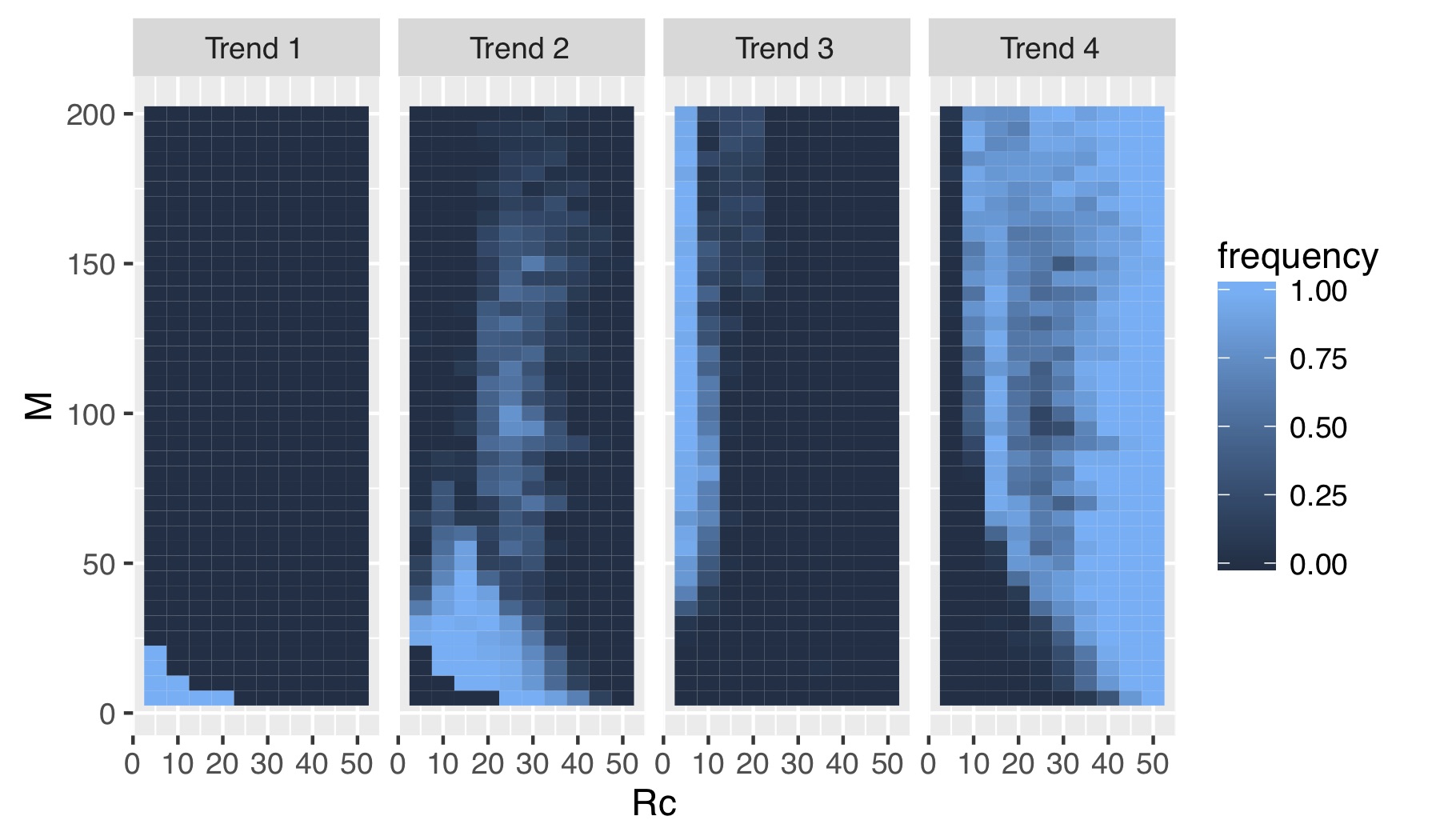}
\caption{
  Frequencies of evolution in each trend.
  Trend 1 evolved when $R_c M$ was small.
  Trend 4 evolved when $R_c M$ was large.
  Trends 2 and 3 evolved in the intermediate range between trends 1 and 4 where $R_c/M$ determined whether groomers evolved to trend 2 or 3.
  Additionally, $R_g$ increased a range of group sizes $M$ in which social grooming strategies evolved to trend 2 or 3, i.e., a large amount of social grooming evolved to trends 2 and 3 in large groups.}
\label{fig_map}
\end{center}
\end{figure}

\begin{figure}[h!]
\begin{center}
\includegraphics[width=0.95\linewidth]{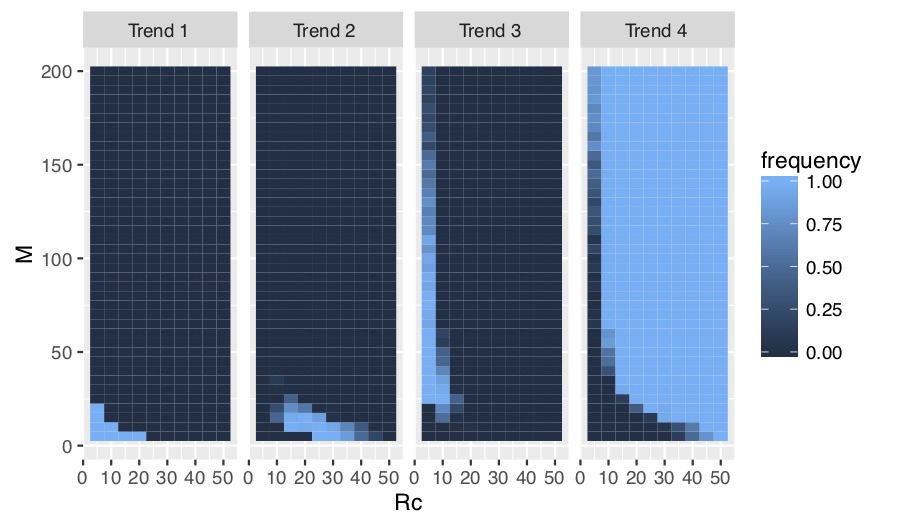}
\caption{The results of evolutionary simulations in $R_g=100, 300$.
  Figures b and d are extended from Figures a and c.
  $q$ increased with $R_c M$.
  This trend in $R_g=100$ was more significant than $R_g=300$.
}
\label{fig_map100}
\end{center}
\end{figure}

We found four evolutionary trends in the results of the simulations (Fig.~\ref{fig_smr}).
These trends are explained by total resources $R_c M$ and the ratios of each cooperator's resources to the number of cooperators $R_c/M$ (Fig.~\ref{fig_map},~\ref{fig_map100}).

\begin{figure}[h!]
\begin{center}
\includegraphics[width=0.95\linewidth]{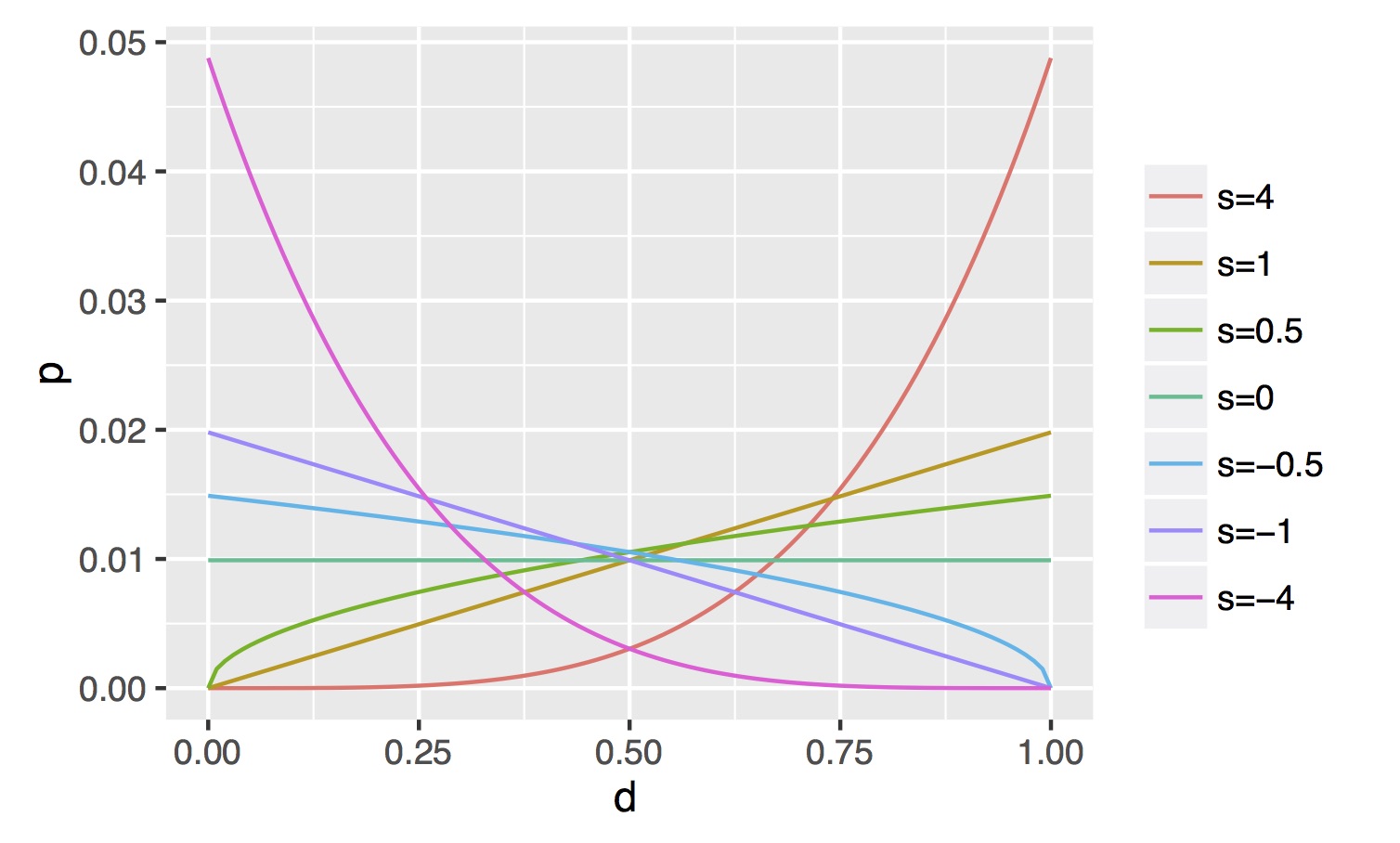}
\caption{
Examples of social grooming strategies.
Social groomers with large $s$ tend to interact with a groomee in a strong social relationship (large $d$).
On the other hand, groomers with small $s$ tend to interact with a groomee in a weak social relationship (small $d$).
When $s=0$, groomers interaction is independent from $d$.
When $s=1$, groomers interact in proportion to the strength of social relationships, i.e. the Yule--Simon process.
}
\label{fig_str_sample}
\end{center}
\end{figure}

Groomers evolved to trend 1 when $R_c M$ was small.
Their $s$ evolved larger and their $q$ evolved smaller.
This strategy concentrates investment into strong social relationships (e.g. $s=4$ in Fig.~\ref{fig_str_sample}).
Groomers tended to evolve to trend 4 when $R_c M$ was large with $s<0$ .
This strategy widely invests in many weak social relationships (e.g. $s=-4$ in Fig.~\ref{fig_str_sample}).
These trends' $s$ do not converge, meaning that they do not have characteristic values.

On the other hand, $s$ converged to $0 < s < 2$ in trends 2 and 3.
Trends 2 and 3 evolved in the intermediate range between trend 1 and 4, and $R_c/M$ determined whether groomers evolved to trend 2 or 3.
Groomers evolved to trend 2 when $R_c/M$ was large, where $q$ evolved larger.
They evolved to trend 3 when $R_c/M$ was small, where $q$ evolved smaller.
$s$ in trend 2 tends to be larger than $s$ in trend 3.
Both strategies are diversified investments (e.g. $s=1$ and $s=0.5$ in Fig.~\ref{fig_str_sample}), where groomers intensively invest in strong social relationships while also widely investing in weak social relationships.
Additionally, $M$, where groomers evolved to trends 2 and 3 is larger, when $R_g$ is large (see Fig.~\ref{fig_map}).

\begin{figure}[h!]
 \begin{center}
 \includegraphics[width=0.95\columnwidth]{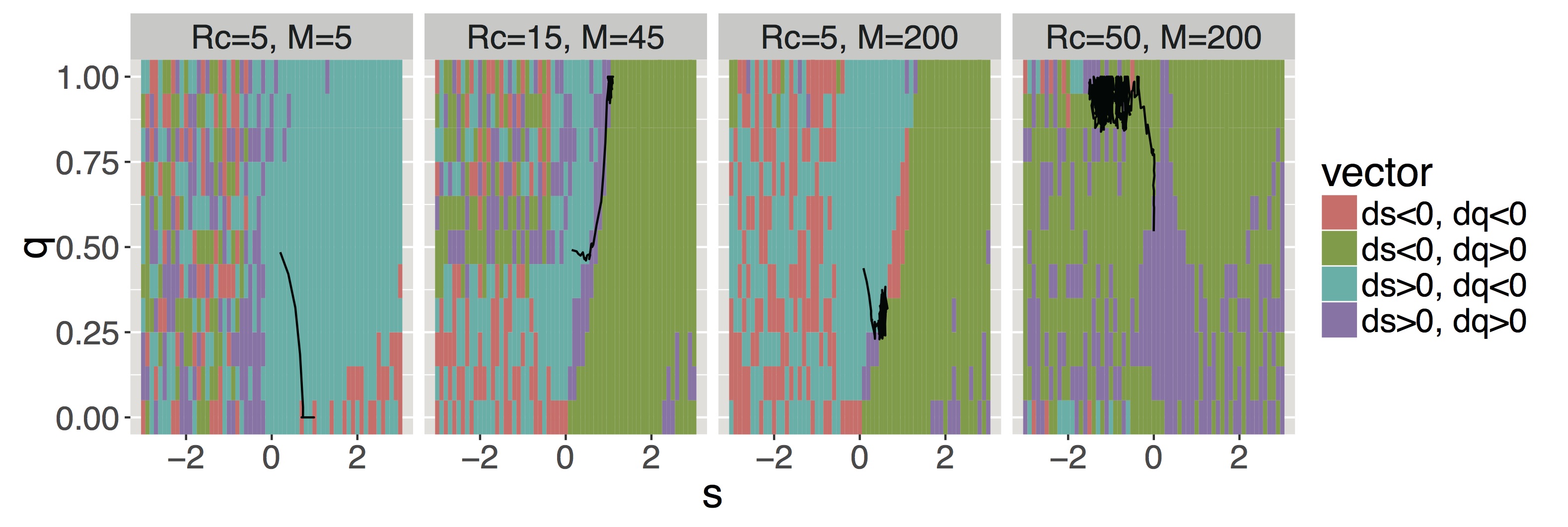}
  \caption{
  Average selection pressures ($ds, dq$) in four trends for $(s, q)$ and typical orbits from $(s, q)=(0, 0.5)$.
  These figures show trends 1, 2, 3, and 4 from left.
  The cell colors show the gradients of $s$ and $q$ (i.e. $ds, dq$).
  For example, the figures show $s$ and $q$ decrease when $(s, q)$ is in a cell of $ds < 0$ and $dq < 0$.
Areas, where the cell colors are mixed, show little gradients, that is, mutation noises were larger than selection pressures.
  For example, populations were random walk along the $s$ axis and they were small along the $q$ axis, when $s < -1$ in $R_c=5$ and $M=200$.
  Evolutionary pressures were calculated using the method of the average gradient of selection (AGoS)~\cite{pinheiro2012}.
  That is, we calculated the mean difference of $s$ and $q$ of the next generation of a population in which individuals' $s$ and $q$ obeyed the Gaussian distribution ($(\mu=s, \sigma=0.2)$ and $(\mu=q, \sigma=0.2)$) on each cell $(s, q)$.
  If the distribution generated a value outside of the range $[0, 1]$ then that was set to the nearest value in the range.
  These orbits were drawn based on the average selection pressures and noises which is a normal distribution with $\mu=0$ and $\sigma=0.01$.
  Incidentally, there is no cell in $(ds, dq)=(0, 0)$.
  Evolutionary dynamics in $R_g=100$ showed similar trends (see Fig.~\ref{fig_vecfield100}).
  }
  \label{fig_vecfield}
 \end{center}
\end{figure}

\begin{figure}[h!]
 \begin{center}
 \includegraphics[width=0.95\columnwidth]{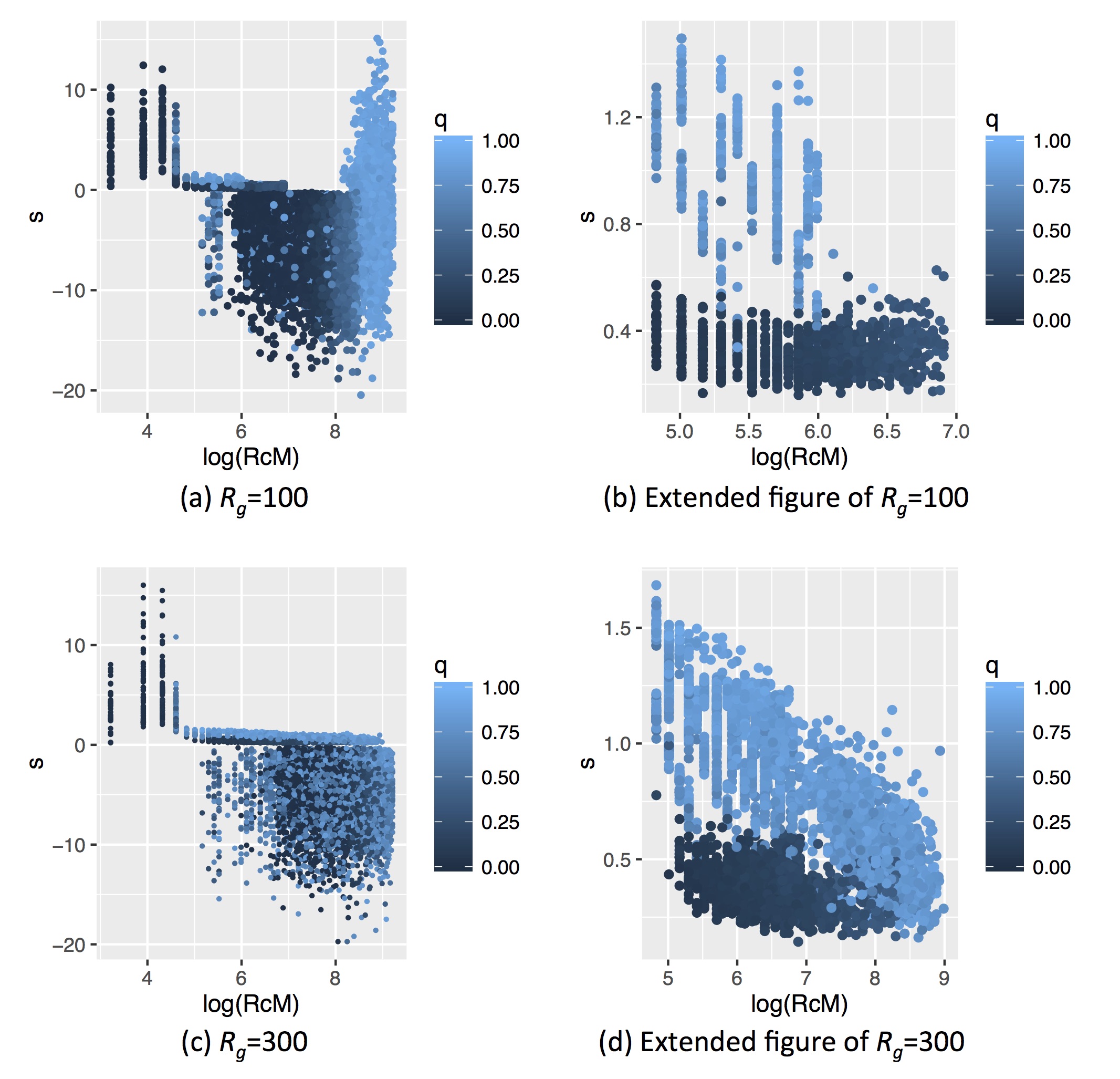}
  \caption{
  Average selection pressures ($ds, dq$) for four trends for $(s, q)$ and typical orbits from $(s, q)=(0, 0.5)$.
  These figures show trends 1, 2, 3, and 4 from left.
  Evolutionary dynamics in $R_g=300$ showed similar trends (Fig. \ref{fig_vecfield}).
  }
  \label{fig_vecfield100}
 \end{center}
\end{figure}

\begin{figure}[h!]
 \begin{center}
   \includegraphics[width=.95\columnwidth]{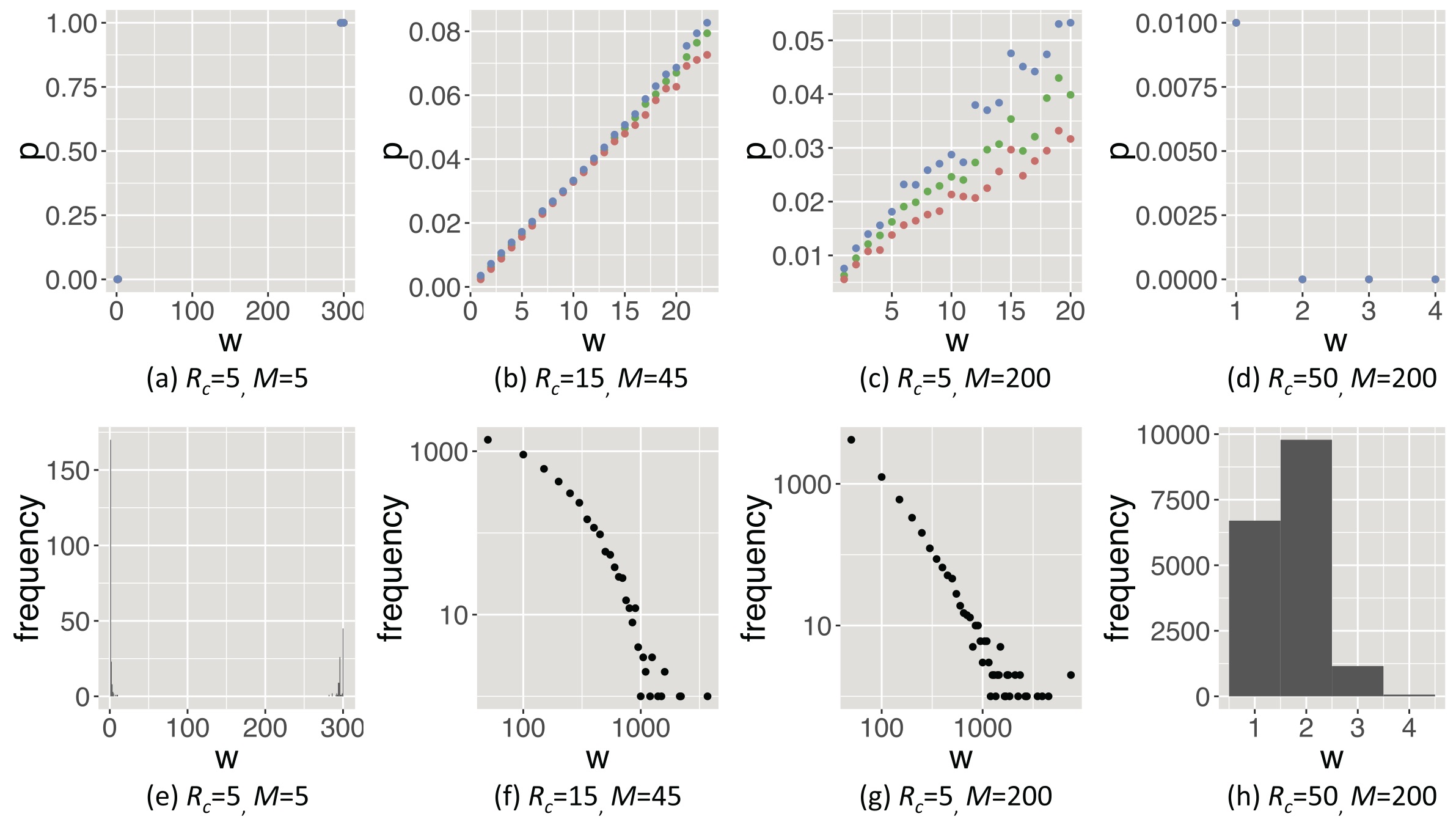}
  \caption{
  Strategies of social grooming (a-d), i.e., probability $p$ of social grooming after each strength of social relationship $w$, and social structures of each trend (e-h), i.e. distribution of $w$ in each trend.
  These figures show trend 1, 2, 3, and 4 from left.
  These trends in $R_g=100$ are similar to them (see S3 Fig).
  In the Figures a-d, the orange points are the $25$th percentile, the green points are the $50$th percentile and the blue points are the $75$th percentile.
  In the Figures a-d, we drew $w$ when the number of samples was more than 20.
  The figures of trend 2 and 3 of the Figures f and g are shown by using a logarithmic scale in both axes.
  In the social structure of trend 1 (e), many weak relationships were caused by mutation noises of $q$.
  }
 \label{fig_w_yule300}
 \end{center}
\end{figure}

\begin{figure}[h!]
 \begin{center}
   \includegraphics[width=.95\columnwidth]{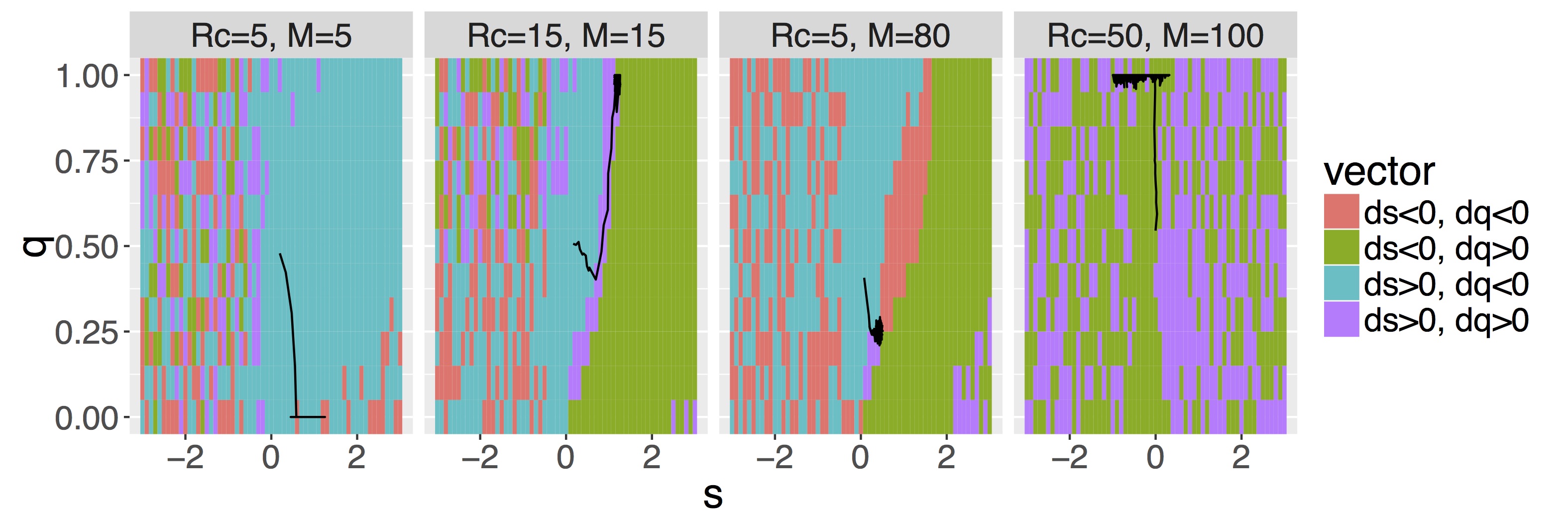}
  \caption{Strategies of social grooming (a-d), i.e., probability $p$ of social grooming after each strength of social relationship $w$, and social structures of each trend (e-h), i.e. distribution of $w$ in each trend.
  These trends in $R_g=300$ are similar to them (see Fig. 6).
  In the Figures a-d, the orange points are the $25$th percentile, the green points are the $50$th percentile and the blue points are the $75$th percentile.
  In the Figures a-d, we drew $w$ when the number of samples was more than 20.
  The figures of trend 2 and 3 of the Figures f and g are shown by using a logarithmic scale in both axes.
  In the social structure of trend 1 (e), many weak relationships were caused by mutation noises of $q$.}
 \label{fig_w_yule100}
 \end{center}
\end{figure}

Next, we demonstrate how the four trends emerged throughout the evolution and how groomers constructed social structures in each trend.
Regarding the former, Fig.~\ref{fig_vecfield}, \ref{fig_vecfield100} shows the evolutionary pressures of each combination of $s$ and $q$, and the typical orbits of evolution.
For the latter, Fig.~\ref{fig_w_yule300}, \ref{fig_w_yule100} shows strategies of social grooming (a-d) and social structures of each trend, i.e. distributions of $w$ (e-h).

Trend 1 evolved in environments with small $R_c M$.
Groomers are in intense competition for receiving cooperation from groomees in the environments.
Therefore, they evolved to concentrate investments to a few poor groomees, i.e. large $s$ and small $q$ ($(R_c, M) = (5, 5)$ in Figs~\ref{fig_vecfield} and \ref{fig_w_yule300}a).
The results show that they only had very strong social relationships in environments with small $R_c M$ (Fig.~\ref{fig_w_yule300}e).

Trend 4 evolved in environments with large $R_c M$.
Groomers easily receive cooperation from groomees in these environments.
Thus, they constructed many weak social relationships with many rich cooperators ($(R_c, M)=(50, 200)$ in Fig.~\ref{fig_vecfield} and Figs~\ref{fig_w_yule300}d and \ref{fig_w_yule300}h)

Trends 2 and 3 evolved between trend 1 and trend 4.
Their $s$ converge to $(0, 2)$, this means groomers with these strategies intensively invest in strong social relationships while they also widely invest in weak social relationships ($(R_c, M)=(15, 45)$ and $(5, 200)$ in Fig.~\ref{fig_vecfield}).
Their social grooming probability is in proportion to each strength of the social relationships (Figs~\ref{fig_w_yule300}b and \ref{fig_w_yule300}c), so their construction processes of social relationships are similar to the Yule--Simon process.
As a result, their social structures were similar to power law distributions (Figs~\ref{fig_w_yule300}f and \ref{fig_w_yule300}g).

\begin{figure}[h!]
 \begin{center}
   \includegraphics[width=.95\columnwidth]{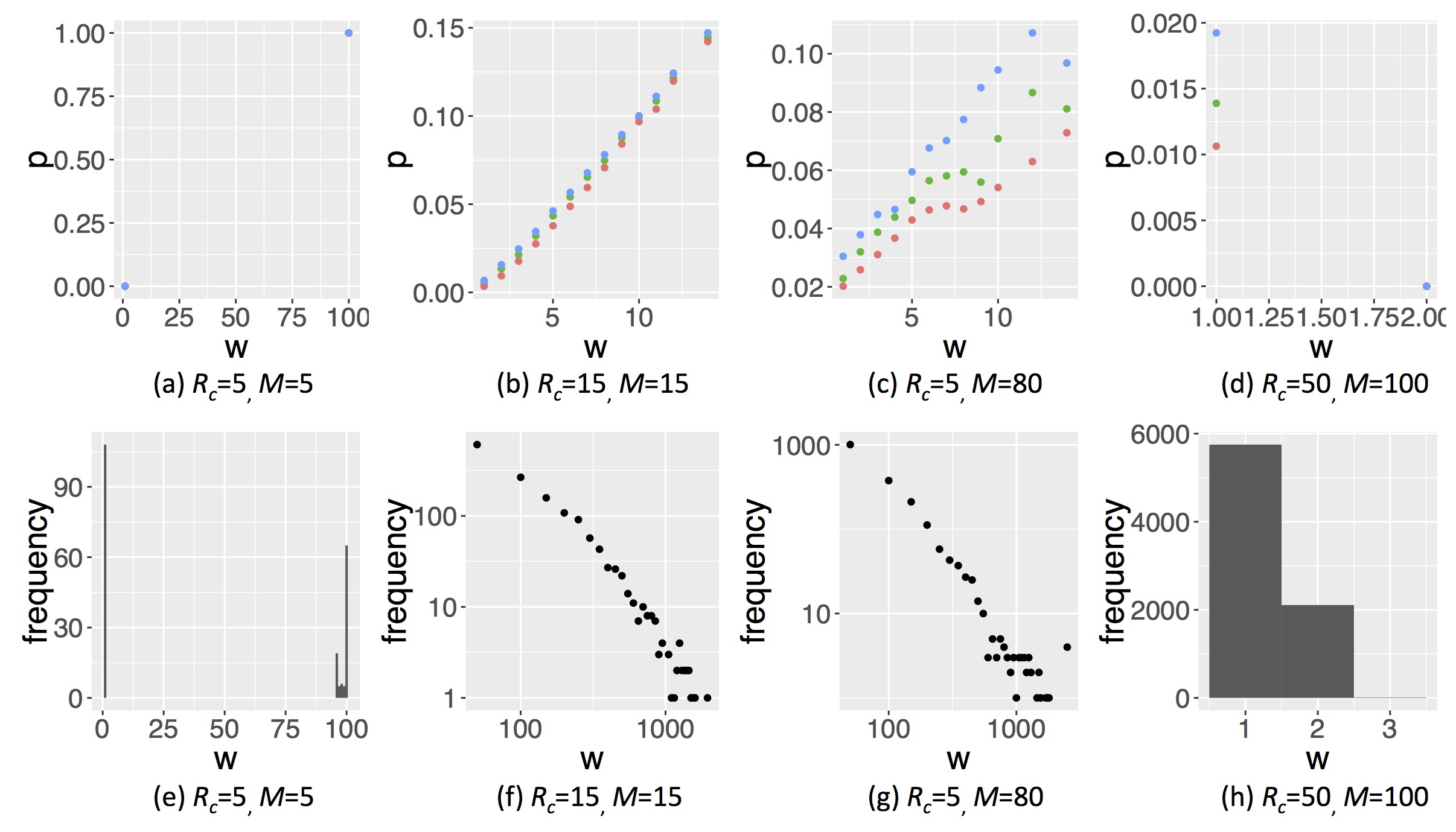}
  \caption{
  Evolution of $q$ in trends 2 and 3 as $\log(R_c/M)$.
  These figures show a non-continuous transition between them.}
 \label{fig_tran}
 \end{center}
\end{figure}

The main difference between trends 2 and 3 is how $q$ is affected by $R_c/M$.
When $R_c/M$ is small, groomers have to confine the number of social relationships with groomees to construct strong social relationships, because they compete intensively in each social relationship (i.e. small $R_c$).
Therefore, they evolved to small $q$ with small $R_c/M$ (trend 3; $(R_c, M)= (5, 200)$ in Fig.~\ref{fig_vecfield}).
In contrast, when $R_c/M$ is large, they do not have to restrict the number of social relationships with groomees, because their competition is not intense in each social relationship (i.e. large $R_c$) and the maximum number of their social relationships is small (i.e. small $M$).
Thus, they evolved to large $q$ with large $R_c/M$ (trend 2; $(R_c, M)= (15, 45)$ in Fig.~\ref{fig_vecfield}).
Interestingly, these trends of evolution show non-continuous transition (see Fig.~\ref{fig_tran}).

\section{Discussion}

We analyzed the evolutionary dynamics of social grooming strategies and social structures.
As a result, we find that the evolutionary dynamics depend on total resources (i.e. $R_c M$) and the ratios of each cooperator's resources to the number of cooperators (i.e. $R_c/M$).
In the poor small groups, individuals' strategies evolved to concentrate investment among strong social relationships.
In the rich large groups, their strategies evolved to wide investment among many weak social relationships.
In the middle groups, their strategies evolved according to the Yule--Simon process.
These strategies invest intensively in strong social relationships while also investing widely in weak social relationships.
As a result of these strategies, skewed distributions of social relationship strengths were generated.

There are two trend strategies which are similar to the Yule--Simon process~\cite{pachur2012,takano2016_palg}.
One evolved in relatively rich and small groups in the middle groups.
Individuals with this strategy constructed social relationships with all group members, and reinforced their relationships in proportion to the strength of social relationships.
The other one evolved in relatively poor and large groups in the middle groups.
Individuals with this strategy constructed social relationships with parts of their groups, and reinforced their relationships.
In primitive human groups, individuals belong to large groups and interact in small cliques within them~\cite{Dunbar2012a}.
Hence, humans' social grooming strategy may have evolved in the latter group.
Non-human primates may also have similar strategies, because they also construct skewed social structures even though their group sizes are different from humans~\cite{Kanngiesser2011,Dunbar2012a,Tung2015,leve2016}.
Their strategies' similarity may be explained by the difference of the amount of social grooming $R_g$.
Our experiments show the increase in the amount of social grooming $R_g$ results in the increase of group sizes $M$, in which social grooming strategies evolve according to the Yule--Simon process (see Fig.~\ref{fig_map}).
The same social grooming strategies are stable in different group sizes.
Actually, there is a positive correlation between group sizes and the amount of social grooming in primates~\cite{dunbar93,dunbar_book2016}.

If a social grooming strategy based on the Yule--Simon process is universal in primates not limited to humans, and group sizes depend on external factors (e.g., predators, food, etc.), then social grooming strategies of humans and non-human primates evolved to the same strategies by automatically adjusting their amount of social grooming.
This relationship between group sizes and strategies may be clearly demonstrated by comparison among humans, non-human primates, and other social animals.
This will contribute towards an explanation of the evolution of humans' large social groups.

It is also important how cooperators select other cooperators as their interaction partners~\cite{hauert2002}.
For example, if cooperators maintain relationships with other cooperators and break relationships with exploiters, their reciprocal relationships will be maintained and their inegalitarian relationships will be broken~\cite{Perc2010,Perc2017}.
This mechanism to keep cooperation is known as network reciprocity.
Social grooming strategies are network construction strategies.
Actually, social grooming has a beneficial effect on the construction of reciprocal relationships~\cite{takano_ngc,takano_socinfo}.
Our results suggest that the evolution of human-like strategies for network construction depends on the resources of environments and their group size.
In this paper, we focused on the evolutionary dynamics of social grooming with stable cooperative behavior.
The co-evolutionary dynamics of both behaviors is an issue to be addressed in the future.

Comparison among various species' data sets will be needed in order to clear the relationships between environments and the four evolutionary scenarios of social grooming strategies.

\section*{Conflict of Interest Statement}

Masanori Takano is an employee of CyberAgent, Inc. There are no patents, products in development or marketed products to declare.

\section*{Author Contributions}

M.T. designed the research. M.T. constructed the model. M.T. performed the simulation. M.T. and G.I. discussed and analyzed the results. M.T. and G.I. wrote the main manuscript text. All authors reviewed the manuscript.

\bibliographystyle{frontiersinSCNS_ENG_HUMS} 



\end{document}